\documentclass[aps,prl,reprint,superscriptaddress]{revtex4-2}
\usepackage{graphicx}

\usepackage{color,amsmath}
\usepackage[normalem]{ulem}
\usepackage{hyperref}
\usepackage{lipsum}
\usepackage{xcolor}
\usepackage{float}

\newcommand{\sio}{\mathrm{SiO_2}}
\newcommand{\dir}[1]{$\left<#1\right>$}
\newcommand{\fac}[1]{$\left\{#1\right\}$}

\begin{document}
\title{Semiconductor Epitaxy in Superconducting Templates}
\author{M.~F.~Ritter}
\affiliation{IBM Research Europe, S\"aumerstrasse 4, 8803 R\"uschlikon, Switzerland}

\author{H.~Schmid}
\affiliation{IBM Research Europe, S\"aumerstrasse 4, 8803 R\"uschlikon, Switzerland}

\author{M.~Sousa}
\affiliation{IBM Research Europe, S\"aumerstrasse 4, 8803 R\"uschlikon, Switzerland}

\author{P.~Staudinger}
\affiliation{IBM Research Europe, S\"aumerstrasse 4, 8803 R\"uschlikon, Switzerland}

\author{D.~Z.~Haxell}
\affiliation{IBM Research Europe, S\"aumerstrasse 4, 8803 R\"uschlikon, Switzerland}

\author{M.~A.~Mueed}
\affiliation{IBM Almaden Research Center, San Jose, California 95120, USA}

\author{B.~Madon}
\affiliation{IBM Almaden Research Center, San Jose, California 95120, USA}

\author{A.~Pushp}
\affiliation{IBM Almaden Research Center, San Jose, California 95120, USA}

\author{H.~Riel}
\affiliation{IBM Research Europe, S\"aumerstrasse 4, 8803 R\"uschlikon, Switzerland}

\author{F.~Nichele}
\email[email: ]{fni@zurich.ibm.com}
\affiliation{IBM Research Europe, S\"aumerstrasse 4, 8803 R\"uschlikon, Switzerland}

\date{\today}

\begin{abstract}
Integration of high quality semiconductor-superconductor devices into scalable and CMOS compatible architectures remains an outstanding challenge, currently hindering their practical implementation. Here, we demonstrate growth of InAs nanowires monolithically integrated on Si inside lateral cavities containing superconducting TiN elements. This technique allows growth of hybrid devices characterized by sharp semiconductor-superconductor interfaces and with alignment along arbitrary crystallographic directions. Electrical characterization at low temperature reveals proximity induced superconductivity in InAs via a transparent interface. 
\end{abstract}
\maketitle

Hybrid semiconductor-superconductor nanostructures are promising candidates for next generation quantum devices as gate-tunable couplers \cite{Makhlin2001, Storcz2003}, superconducting qubits \cite{Larsen2015, DeLange2015}, Andreev qubits \cite{Zazunov2003, Chtchelkatchev2003, Lee2014, Janvier2015} and qubits based on Majorana zero modes \cite{Lutchyn2010, Oreg2010}. 
Their applications rely on highly transparent semiconductor-superconductor interfaces, a milestone first achieved by the epitaxial growth of Al on InAs nanowires (NWs) \cite{Chang2015, Krogstrup2015} and later on 2D electron gases \cite{Shabani2016, Kjaergaard2016}. In recent pioneering approaches, large gap elemental superconductors such as Nb, Ta, V, Sn and Pb \cite{Bjergfelt2019, Khan2020, Carrad2020, Pendharkar2021, Kanne2021} were coupled to semiconductor NWs via transparent interfaces. These approaches allow exquisite control of the hybrid interface and are compatible with elaborate shadow epitaxy techniques \cite{Gazibegovic2017} but they are challenging to scale and difficult to integrate in a CMOS architecture.\\
Here we demonstrate a different approach in which the order of semiconductor epitaxy and superconductor deposition is reversed. A crystalline semiconductor is grown inside an insulating $ \sio $ template cavity which features integrated superconducting elements, resulting in flat semiconductor-superconductor hybrid interfaces. This technique is scalable and CMOS compatible, as it is based on the template-assisted selective epitaxy (TASE) platform \cite{Schmid2015a, Borg2015, Knoedler2017, Borg2019} where III-V semiconductors are grown inside insulating cavities. In recent years the TASE approach enabled dense integration of III-V nanowires on Si \cite{Schmid2015a}, growth in branched geometries \cite{Gooth2017a} and ballistic transport over hundreds of nanometers \cite{Gooth2017}.  Since our approach involves formation of a superconductor-semiconductor interface, we refer to it here as hybrid-TASE. \\
In this work we introduce hybrid-TASE with InAs NWs and the superconductor TiN. Nanowires are aligned laterally on the substrate and grown along arbitrary crystallographic directions. We investigate the hybrid interface by scanning transmission electron microscopy (STEM) and present a detailed study of InAs epitaxy in various templates. Finally, we present tunneling spectroscopy of a proximitized hybrid-TASE NW.

\begin{figure*}
	\includegraphics[width=2\columnwidth]{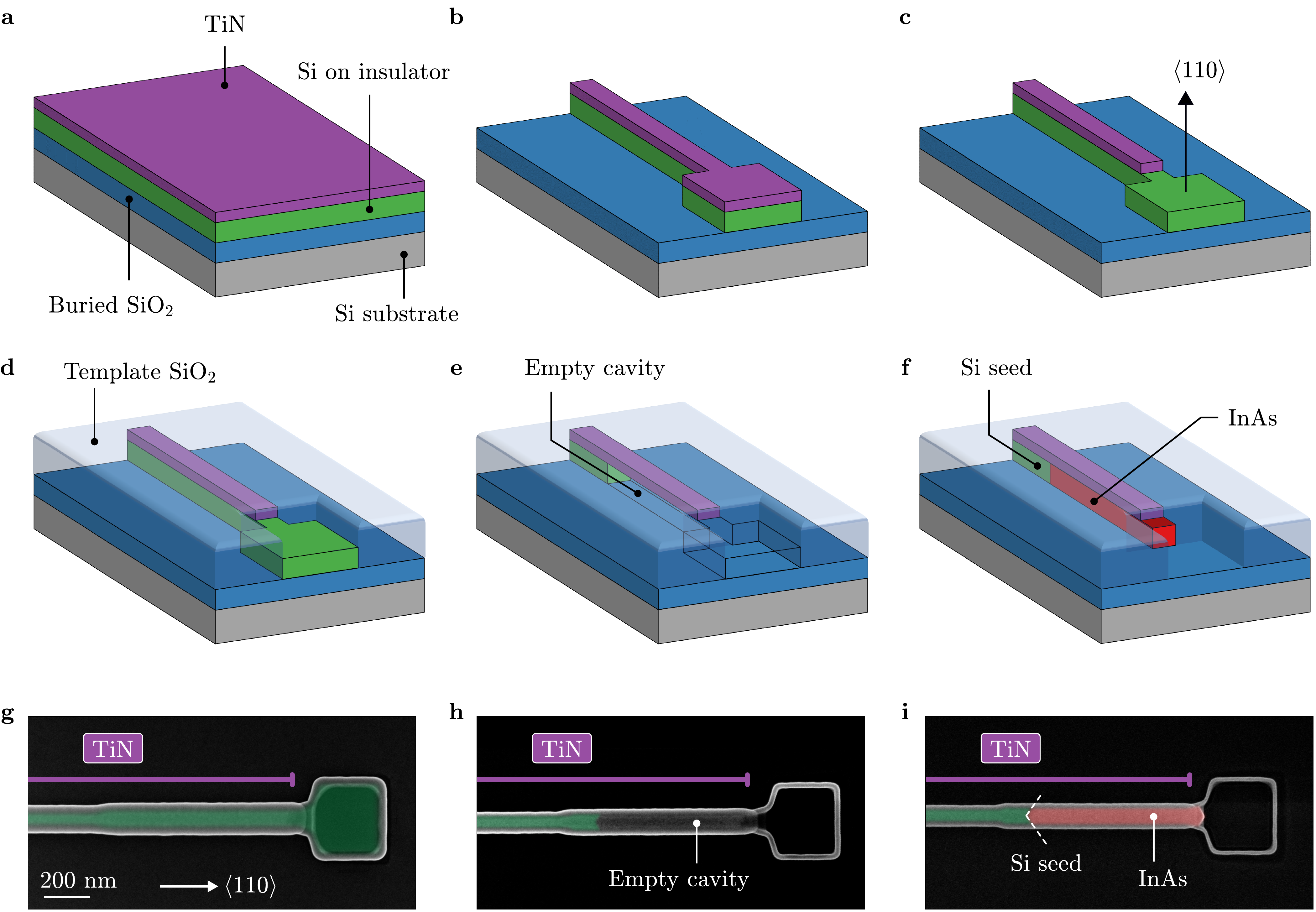}
	\caption{\textbf{Deterministic hybrid-TASE nanowire growth inside a lateral template.}
		\textbf{a} A silicon-on-insulator wafer consisting of a thin Si layer (green) separated from the Si substrate (gray) by a buried $ \sio $ layer (blue) is metallized with a 25~nm thick film of TiN. \textbf{b} Self-aligned TiN/SOI bilayer structures are patterned. The structure terminates in a square. \textbf{c} TiN is etched from this square. \textbf{d} Devices are covered in a conformal 40~nm template $ \sio $ layer (light blue). $ \sio $ is locally etched to expose the Si square at the wire end. The TiN stripe remains protected. \textbf{e} Selective etching of the SOI creates a cavity formed by template $ \sio $ and the TiN stripe. A segment of Si remains at the end of the cavity. \textbf{f} The surface of the Si segment acts as a seed for epitaxial growth of InAs nanowires (red). InAs nanowires are guided by the template cavity and an interface to TiN is formed during InAs epitaxy \textbf{g,h,i} Top-view SEM micrographs of the fabrication steps in subfigures \textbf{d} ,\textbf{e} and \textbf{f}, respectively. Regions of Si (green) and InAs (red) are false colored, they are located below the TiN stripe and $ \sio $ template layer.  The extent of the TiN stripes integrated into the template is indicated by purple lines. Dashed lines in \textbf{i} indicate Si \fac{111} seed facets.}
	\label{fig:1}
\end{figure*}


We based the fabrication of hybrid-TASE templates on commercial Si wafers that featured a 150~nm thick buried-oxide (BOX) layer and a 40~nm or 70~nm thin crystalline silicon-on-insulator (SOI) top layer. The SOI layer had a $ \left(110\right) $ orientation, different from the $ \left(001\right) $ SOI layers used in previous work \cite{Schmid2015a, Borg2015, Knoedler2017, Gooth2017, Gooth2017a,Borg2019}. This allowed us to laterally integrate III-V nanostructures along different directions on a single chip, such as \dir{100}, \dir{110}, \dir{111}, \dir{112} and even lower symmetry directions. Our choice of TiN as the superconductor is motivated by its chemical inertness and refractory properties, making it compatible with our process flow and the high temperatures reached during semiconductor epitaxy.\\
Figures~\ref{fig:1}a-f show a simplified schematics of the hybrid-TASE process flow, while Figs.~\ref{fig:1}g-i show scanning electron microscope (SEM) images of a typical device at three fabrication steps, respectively. A detailed description of the process flow is reported in the Methods Section. In the first step, the SOI layer was metallized by sputtering of a 25~nm thick layer of TiN (Fig.~\ref{fig:1}a). Self-aligned TiN/SOI bilayer nanowires were dry etched in a single step (Fig.~\ref{fig:1}b) and TiN was locally wet etched from one end of the wire (Fig.~\ref{fig:1}c), leaving the underlying SOI unaffected. The patterned structures were covered in a conformal 40~nm thick $\sio$ template, which was locally etched at the template termination (Fig.~\ref{fig:1}d and g). Selective wet etching of the SOI resulted in hollow cavities with sidewalls of $ \sio $, the BOX layer as floor and TiN as ceiling. The length of the cavity was determined by the SOI etching time (Fig.~\ref{fig:1}e and h). Cavities formed in this way terminated in a crystalline Si surface originating from the original SOI layer, serving as a nucleation seed for InAs heteroepitaxy. InAs nanowires were grown inside the template structures via metal-organic chemical vapor phase epitaxy (MOVPE) (Fig.~\ref{fig:1}f and i) using trimethylindium (TMIn) and tertbutylarsine (TBAs) as precursor species. 
Height and width of the resulting InAs nanowires were determined by the SOI layer thickness and template width, respectively. The NW length was determined by the cavity length and growth time. 


\begin{figure*}
	\includegraphics[width=2\columnwidth]{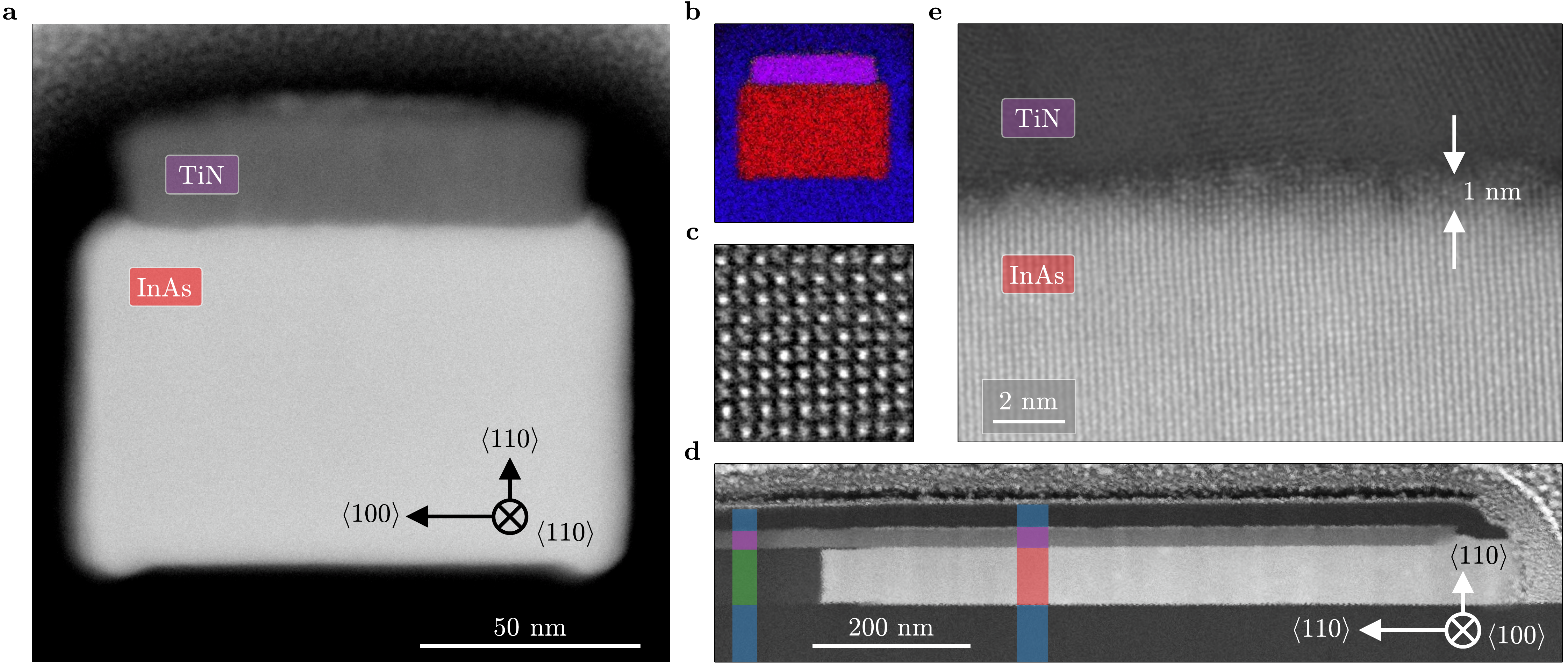}
	\caption{\textbf{Structural study of hybrid-TASE nanowires and the InAs/TiN hybrid interface using STEM.}
		\textbf{a}~Transversal cross section of a hybrid-TASE nanowire similar to that of Fig.~\ref{fig:1}i. The \dir{110} aligned InAs nanowire features a rectangular cross section and is grown inside a hybrid template formed of TiN and $ \sio $. \textbf{b} EDX elemental mapping of the cross section shown in \textbf{a} with In (red), Ti (purple) and Si (blue). The frame size is 160~nm $ \times $ 160~nm. \textbf{c} High-resolution STEM image of InAs along the $ \left<100\right> $ zone axis. The frame size is 2.3~nm $ \times $ 2.3~nm. \textbf{d} Overview of the InAs crystal of \textbf{c} imaged along the $ \left<100\right> $ zone axis. The cut is performed along the axis of the device in Fig.~\ref{fig:3}e. The Si seed and the InAs nanowire form an interface at the InAs nucleation site, a stripe of TiN covers both materials. Colored boxes indicate $ \sio $ (blue), Si (green), InAs (red) and TiN (purple), respectively \textbf{e} Typical zoom-in on the hybrid interface formed by TiN and InAs. The interface roughness between polycrystalline TiN and the InAs single-crystal is approximately 1~nm.} 
	\label{fig:2}
\end{figure*}

\begin{figure}
	\includegraphics[width=\columnwidth]{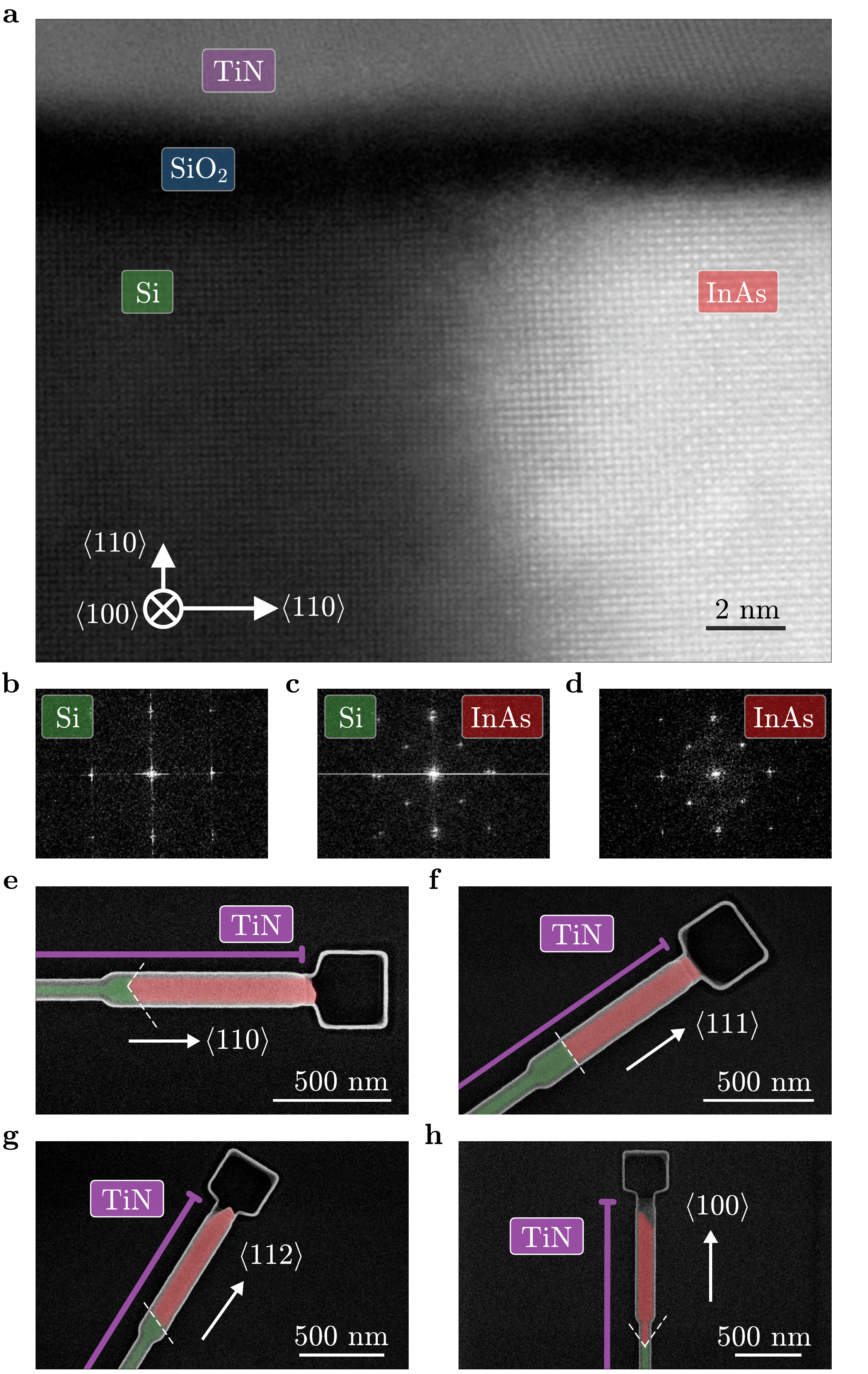}
	\caption{\textbf{Nucleation and growth uniformity of InAs hybrid-TASE.}
		\textbf{a} Detailed view of the seed area of Fig.~\ref{fig:2}a imaged along the $ \left<100\right> $ zone axis. \fac{111} facets of the seed are inclined with respect to the zone axis (see text). \textbf{b, c, d} Fast Fourier transforms of the Si seed layer, the heteroepitaxial interface of Si and InAs and the InAs NW, respectively. \textbf{b} and \textbf{d} highlight the single crystaline structure of Si seed and hybrid-TASE grown InAs, double spots in \textbf{c} are testament to different lattice constants of the materials. \textbf{e} Top-view false color SEM micrograph of a hybrid-TASE NW oriented along \dir{110} direction. The entire length of the InAs wire is covered by a TiN stripe (purple line). \textbf{f, g, h} As in \textbf{e} but with templates oriented along \dir{111}, \dir{112} and \dir{100} direction, respectively. Si \fac{111} seed facets are indicated by dashed lines.}
	\label{fig:3}
\end{figure}

We investigated the structural quality of our nanowires by high-resolution scanning transmission electron microscopy (STEM) and energy-dispersive X-ray spectroscopy (EDX) at 200~kV on lamellae prepared across and along the nanowire axis via focused ion beam techniques. Figure~\ref{fig:2}a presents an annular dark field (ADF)-STEM cross sectional view of a hybrid-TASE nanowire similar to the device shown in Fig~\ref{fig:1}i. The InAs crystal (bright) exhibits a rectangular cross section with a flat interface to TiN \footnote{We attribute the slightly rounded corners and the expansion of the InAs crystal beyond the width of the TiN segment to an enlargement of the $ \sio $ template during a final HF etch prior to InAs epitaxy (see Methods Section).}. EDX elemental mapping shown in Fig.~\ref{fig:2}b highlights the elemental distribution of the templated nanowire with In (red), Ti (purple) and Si (blue). A representative high magnification ADF-STEM image of the InAs crystal along the \dir{100} zone axis (Fig.~\ref{fig:2}c) confirms the single crystalline nature of the NW. Figure~\ref{fig:2}d shows the view along the $ \left<100\right> $ zone axis, obtained by cutting the device illustrated by the SEM image in Fig.~\ref{fig:3}e. Colored boxes in Fig.~\ref{fig:2}d highlight the material stack with $ \sio $ (blue), Si (green), InAs (red) and TiN (purple), respectively. The Si seed is visible on the left-hand-side of Fig.~\ref{fig:2}d as a vertical interface. Due to its \fac{111} facet morphology \cite{Schmid2015a}, the seeds appeared V-shaped when viewed from the [110] direction (Fig~\ref{fig:1}i). This three-dimensional morphology cannot be captured from the $ \left<100\right> $ view presented in Fig~\ref{fig:2}d.\\
We investigated the interface between TiN and InAs by recording high-resolution ADT-STEM images of this region along the growth axis. A typical example is shown in Fig.~\ref{fig:2}e. The interface between InAs and polycrystalline TiN is sharp and free of contaminations, as evidenced by the interface roughness on the order of 1~nm and by EDX elemental line profiles (see \href{}{Supporting Information}). A lower bound for the roughness of the hybrid interface was set by the initial SOI layer roughness of 0.3~nm rms, which is likely to increase during processing prior to TiN deposition (see Methods Section). We also point out that the finite lamella thickness of $ \sim80~\textrm{nm} $ might cause the observed interface roughness to appear larger. In previous studies, the roughness of metal top surfaces was found to promote detrimental parasitic nucleation during semiconductor epitaxy \cite{Kobayashi2013}. In contrast, our approach utilizes the pristine and freshly exposed TiN back surface, which allows selective growth in geometries with high aspect ratio.\\


The morphology of the Si seed from which III-V epitaxy started is detailed in Fig.~\ref{fig:3}. Figure~\ref{fig:3}a shows a high-resolution ADF-STEM image of the interface between Si and InAs from the device presented in Fig.~\ref{fig:2}d. Between Si/InAs and TiN appears a layer of $ \sio $, an artifact of the projection of the three dimensional, V-shaped Si \fac{111} seed facets (Fig~\ref{fig:1}i) onto the viewing plane. The single-crystalline nature is evidence by fast Fourier transforms (FFTs) of the Si seed (Fig.~\ref{fig:3}b), the Si/InAs heterointerface (Fig.~\ref{fig:3}c) and the InAs nanowire (Fig.~\ref{fig:3}d) along the \dir{100} zone axis \footnote{The FFTs are computed from an overview image of the seed area larger than the frame shown in Fig.~\ref{fig:3}a.}. The analysis shows a clear transition from the diamond cubic crystal structure of Si to the zinc blende crystal structure of InAs. The mismatch in lattice constant between Si and InAs is resolved in Fig.~\ref{fig:3}c as double peaks in the FFT. The alignment between the two peaks demonstrates the epitaxial relation between the two materials, confirming InAs nucleated from Si and not from TiN. Detailed comparison of the alignment of the SOI and InAs crystal revealed a rotation of $\sim1^{\circ}$ along the \dir{100} zone axis. This is expected in TASE epitaxy where rotations of up to $ 3^{\circ} $ are observed \cite{Knoedler2017}. EDX data of the seed interface is presented in the \href{}{Supporting Information}. \\

The key concept of hybrid-TASE epitaxy, that is the formation of a hybrid interface during semiconductor growth, required the InAs crystal to radially expand to the template walls. 
We achieved this using a high V/III precursor ratio of 150 and a nominal temperature of $ 550~^{\circ}\mathrm{C} $ to promote growth of \fac{110} facets deep inside cavities, where the local V/III ratio was reduced due to differing diffusion mechanisms of the precursor species \cite{Borg2015}. Furthermore, these conditions enabled an isotropic growth rate along a plethora of orientations. We present devices grown in hybrid templates along the \dir{110}, \dir{111}, \dir{112} and \dir{100} direction in the false colored SEM micrographs of Figs.~\ref{fig:3}e-h. Further examples along lower symmetry directions, which are challenging to grow using standard approaches \cite{Lee2019}, are shown in the \href{}{Supporting Information}. Such NWs are interesting for the investigation of spin-orbit interaction along low-symmetry crystallographic directions \cite{Sasaki2014, Manchon2015, Carballido2021}.  Independently of the wire direction, all devices featured \fac{111} seed facets which were oriented perpendicular to the wafer surface, while their alignment with respect to the InAs NW axis changed depending on the template orientation. The formation of \fac{111} facets is a consequence of the anisotropic Si wet etch, which favors the formation of \fac{111} facets.  In the particular case of \dir{111} aligned templates, this resulted in a single seed and growth facet, perpendicular to both the wafer plane and the wire axis (Fig.~\ref{fig:3}f). Finally, we observed that the presence of a TiN layer impacted the growth dynamics of our nanowires. In particular, hybrid-TASE nanowires displayed an axial growth rate which was up to 4 times higher than that of wires grown with the standard TASE method.
Further epitaxy experiments at decreased precursor flow indicated an increased V/III ratio inside hybrid-TASE templates compared to TASE, likely due to enhanced surface diffusion of the precursors on the flat and inert TiN surface. In the \href{}{Supporting Information} we provide a detailed discussion of the altered growth dynamics in hybrid templates.\\


\begin{figure}
	\includegraphics[width=\columnwidth]{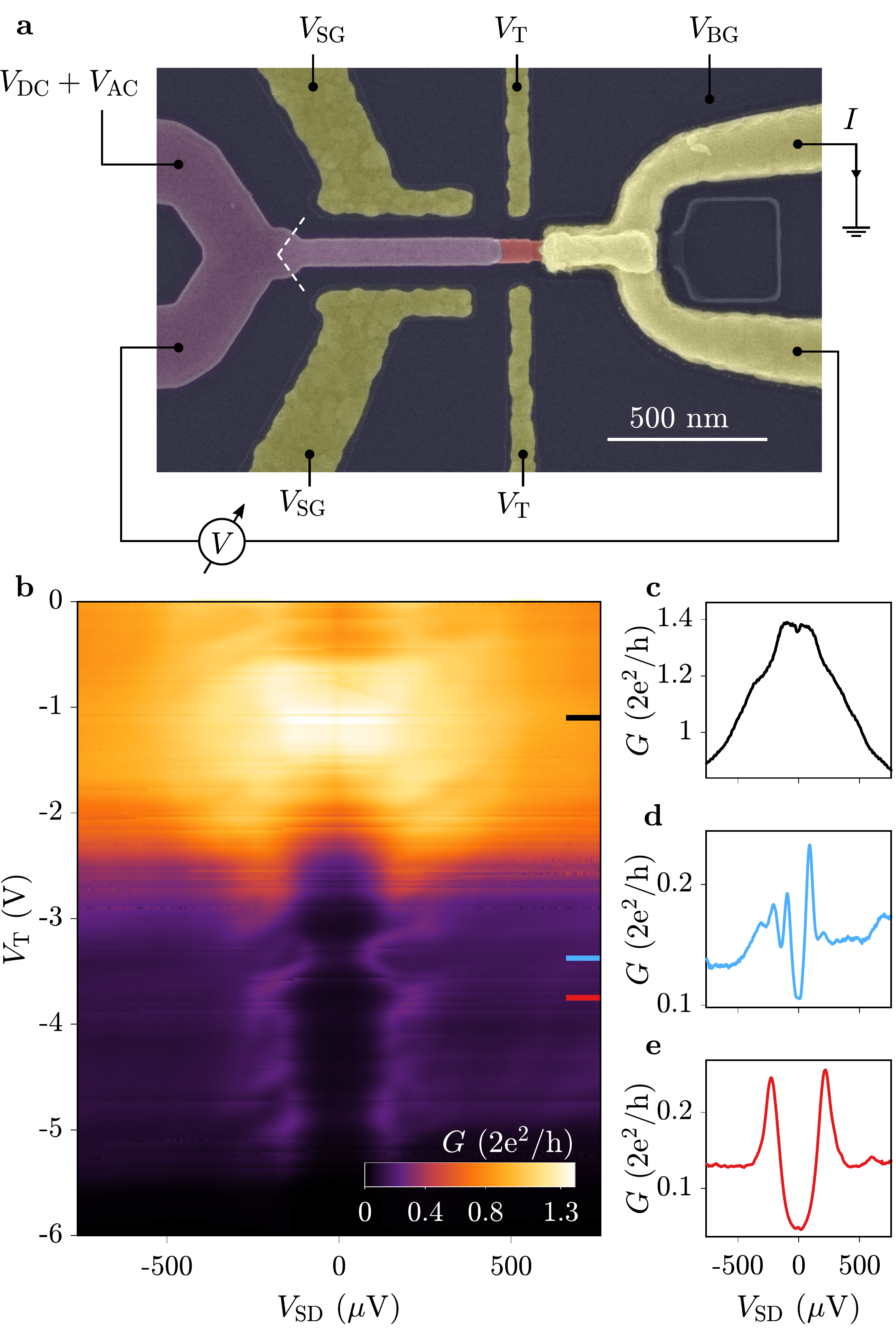}
	\caption{\textbf{Finite bias spectroscopy of a hybrid-TASE tunnel junction.}
		\textbf{a} False colored SEM micrograph of a hybrid-TASE device where InAs (red) is grown inside a template with integrated TiN segment and contacts on one side (purple). Normal contacts and gates (yellow) are patterned with lift-off. Dashed lines indicate the InAs nucleation site below TiN. \textbf{b} Finite bias spectroscopy of a hybrid-TASE tunnel contact formed by tuning the side gate voltage $ V_{\textsc{T}} $ at $ V_{\textsc{BG}}=-15~\textrm{V} $ and $ V_{\textsc{SG}}=0~\textrm{V} $. \textbf{c, d, e} Line cuts of the data in $ \textbf{b} $ at $V_{\textsc{T}}=-1.1~\textrm{V} $, $ V_{\textsc{T}}=-3.375~\textrm{V} $ and $V_{\textsc{T}}=-3.75~\textrm{V} $, respectively.}
	\label{fig:4}
\end{figure}

We performed electrical characterization of the hybrid TiN/InAs interface by means of finite bias spectroscopy on the device of Fig.~\ref{fig:4}a. It featured a total length of 1.16~$ \mu $m and a cross section of 50~nm $ \times $ 80~nm. We altered step 3 of the hybrid-TASE process flow (Fig.~\ref{fig:1}c) such that TiN was etched on a 560~nm long segment, allowing for a normal metal probe to be integrated after InAs growth. We also patterned side gates on either side of the wire. Both the normal contact and the gates (yellow in Fig.~\ref{fig:4}a) were fabricated by evaporation of Ti/Au and lift-off. On the seed side of the wire, the TiN layer branched off to bonding pads \footnote{This specific device was grown with a precursor ratio of V/III=70.}. The Si substrate (Fig.~\ref{fig:1}a) was metallized on the backside by evaporation of Ti/Pt and used as a global back gate. \\
Measurements were performed by low-frequency lock-in technique at a temperature of 20~mK. A voltage bias $V_{\mathrm{DC}}+V_{\mathrm{AC}}$ was applied at one end of the nanowire while the resulting voltage difference $V$ and current to ground $I$ were measured via a differential voltage amplifier and a low impedance IV converter, respectively.
As the Si handle-wafer became insulating below 10~K, we used the back-gate voltage $V_{\mathrm{BG}}$ to define the operating point of the device at 14~K (see \href{}{Supporting Information}) and subsequently cooled down the device to mK temperatures. Devices prepared in this way showed remarkable electrical stability over several days of measurements.\\
Data presented in Fig.~\ref{fig:4}b was obtained by applying a voltage $ V_{\textsc{T}} $ to two side gates (see Fig.~\ref{fig:4}a), and recording the differential conductance $G$ as a function of the source-drain voltage $ V_{\textsc{SD}} $. Three distinct regimes are identified, based on the normal state transmission of the tunneling probe (see linecuts in Figs.~\ref{fig:4}c-e). At $ V_{\textsc{T}} =-1.1 ~\textrm{V}$ the conductance at small $ V_{\textsc{SD}} $ was enhanced, a hallmark of Andreev reflection (Fig.~\ref{fig:4}c). The conductance spectrum at $ V_{\textsc{T}} =–3.375 ~\textrm{V}$ highlights discrete subgap states (Fig.~\ref{fig:4}d). Further decreasing the transmission, at $ V_{\textsc{T}} =– 3.75 ~\textrm{V}$, we measured an induced superconducting gap of $ \Delta^{\ast} = 220~\mu\textrm{eV}$ (Fig.~\ref{fig:4}e). These findings are consistent with a highly transparent semiconductor-superconductor interface, with electronic transport governed by Andreev reflection \cite{Beenakker1992, Chang2015,Kjaergaard2016}.
The induced superconducting gap $ \Delta^{\ast} $ depended on the specific gate tuning and we achieved the highest value of $ \Delta^{\ast} = 300~\mu\mathrm{eV} $ at $ V_{\mathrm{BG}}=0~\mathrm{V} $ and $ V_{\mathrm{SG}}=-5~\mathrm{V} $. For bulk TiN the expected superconducting gap is $ \Delta = 500~\mu\textrm{eV}$ \cite{Pracht2012}. A possible cause for the reduced superconducting gap in our device is degradation of TiN during fabrication. Measurements on reference TiN nanowires highlighted a decrease of the TiN critical temperature from $3.7$ to $3.0~\textrm{K}$ after encapsulation in $ \sio $ and annealing at 600~$ ^{\circ}\textrm{C} $ for 30~s. In contrast, $ T_{\textrm{C}} $ of TiN wires encapsulated in a SiN$ _{\textrm{x}} $ template remained unchanged after annealing for 30~minutes. To avoid degradation of the superconductor during high temperature processing, future devices could, therefore, use a SiN$ _{\textrm{x}} $ template dielectric, as routinely employed in selective-area grown devices \cite{Het2020}.\\\\
Our approach to semiconductor-superconductor device fabrication is complementary to existing methods, which are based on the in-situ growth of elemental superconductors on semiconductors at low temperatures. Furthermore, hybrid-TASE enables new semiconductor-superconductor material combinations. In particular, TASE was already demonstrated for semiconductors such as GaAs, InSb and GaSb \cite{Schmid2015a, Knoedler2017, Borg2015}. Future hybrid devices might employ superconductors which are chemically similar to TiN, but characterized by higher critical temperatures and magnetic fields, such as NbN and VN, making the hybrid-TASE platform particularly interesting for applications requiring high magnetic fields. The compatibility of hybrid-TASE with standard CMOS fabrication can furthermore enable 3D integrated \cite{Schmid2015a} cryogenic qubit control electronics at few K temperature such as amplifiers and multiplexers with low power dissipation, beyond the offerings of Si CMOS \cite{Zota2019}.\\\\
We presented epitaxy of InAs nanowires on Si inside superconducting TiN/$ \sio $ lateral cavities, a scalable and CMOS compatible approach to semiconductor-superconductor hybrids. We demonstrated InAs growth in a large variety of crystal directions and observed enhanced growth rates in the presence of exposed TiN. Transport spectroscopy revealed proximity induced superconductivity in the semiconductor, with a transparent hybrid interface. 
\section{Methods}
\textbf{Marker Fabrication.} Before patterning the hybrid-TASE templates, we defined markers for optical and electron beam lithography. First, we deposited a 30~nm layer of $ \sio $ via plasma-enhanced chemical vapor deposition, then a 100~nm layer of W via sputtering. Using electron beam lithography, we exposed markers on a AR-N 7520.17 negative tone and transferred the pattern into W by reactive ion etching (RIE) in N$ _2 $/SF$ _6 $ plasma, using the $ \sio $ layer as etch stop. After removing the resist, we encapsulated the markers in 300~nm $ \sio $ grown with plasma-enhanced chemical vapor deposition using tetraethyl orthosilicate as precursor. The wafers were annealed at 750~$ ^{\circ} $C for 30~s and device areas were defined via optical lithography and buffered hydrofluoric acid (BHF) etching by exposing the SOI layer in regions where hybrid-TASE templates will be patterned.\\
\textbf{Fabrication of templates with integrated TiN segments.}
Wafers were cleaned in concentrated piranha solution (sulfuric acid and hydrogen peroxide 2:1) followed by a rinse in ultrapure water and cleaning in a 600~W oxygen plasma. The native $ \sio $ layer was thick enough to protect the back face of the TiN layer during wet etching of Si, greatly enhancing the fabrication yield.\\
Subsequently, we sputtered a 25~nm thick layer of TiN on the entire wafer via DC reactive magnetron sputtering (Fig.~\ref{fig:1}a). \\
We patterned Si/TiN bilayer nanostructures via inductively coupled HBr plasma etching. For this purpose we defined a 50~nm thick layer of hydrogen silsesquioxane (HSQ) negative tone resist as etch mask, using electron-beam lithography. After etching, HSQ was removed in diluted hydrofluoric acid. Typical Si/TiN wires patterned in this fashion were 2~$ \mu $m long and had a width ranging from 40~nm to 100~nm. The lithographically defined width corresponds to the width of InAs nanowires grown inside hybrid-TASE templates. Si/TiN wires terminated in a square, as shown in Fig.~\ref{fig:1}b.\\
To ensure adhesion of a 80~nm AR-P 6200.04 positive tone resist layer on TiN, we encapsulated the structures in a 5~nm $ \sio $ layer deposited via oxygen plasma ALD. Using electron-beam lithography, we defined regions for TiN etching and etched the exposed ALD grown $ \sio $ layer in BHF. We selectively removed TiN in a wet-etch solution of $ \mathrm{H_{2}O}$, $\mathrm{H_{2}O_{2}}$ and $ \mathrm{NH_{4}OH} $ (5:2:1) at 65~$ ^{\circ}\mathrm{C} $ \cite{Heo2012} as indicated in Fig.~\ref{fig:1}(c). \\
The resist was removed with organic solvents and a 40~nm layer of $ \sio $ was deposited using oxygen plasma enhanced ALD. This $ \sio $ layer will guide the growth of III-V structures and we refer to it as $ \sio $ template. To reduce the template etch rate in diluted HF, we annealed devices at 600~$ ^{\circ} $C for 30~s in Ar/H$ _{2} $ atmosphere. Using electron beam lithography on a 80~nm layer of AR-P 6200.04 positive tone resist, openings in areas where TiN had been etched previously were defined. We transferred the openings into the $ \sio $ template using RIE in Ar/$ \mathrm{CHF_{3}} $ plasma and BHF etching. In this way, the Si square at the end of each wire was exposed. Importantly, the exposed area did not overlap with TiN segments on top of the sacrificial Si wire, i.e. TiN features remained protected by $ \sio $. This situation is sketched in Fig.~\ref{fig:1}d.\\
The exposed Si square allowed us to selectively etch the sacrificial Si structures, creating cavities formed by template $ \sio $ and TiN. We etched Si in a 2~\% tetramethylammonium hydroxide solution at 80~$ ^{\circ}$C. The cavity length was determined by the etching time which was chosen such that a segment of Si remained, as shown in Fig.~\ref{fig:1}e. Due to the anisotropy of the etch, residual Si segments exhibited typical $ \left\{ 111 \right\} $ facets as seen in Fig.~\ref{fig:1}h. The facets were oriented perpendicular to the (110) wafer surface.
\\
\textbf{InAs epitaxy inside hybrid-TASE template cavities.}
Prior to MOVPE semiconductor growth, we immersed the templates in diluted hydrofluoric acid H$ _{2} $O:HF 20:1. The etching served two purposes as it removed both the native $ \sio $ protection layer below the TiN segments and etched native $ \sio $ from the Si $ \left\{111\right\} $ seed facets while creating hydrogen terminated facets. At the same time, the inner template dimensions increased slightly. This effect can be seen in Fig.~\ref{fig:2}a, where the InAs nanowire was approximately 20~nm wider than the TiN region.\\
We promptly transferred chips into a MOVPE growth reactor where they were annealed at 600~$ ^{\circ} $C for 5~min under TBAs flow. $ \mathrm{H}_{2} $ was used as carrier gas and InAs growth started as we introduced TMIn into the reactor. InAs growth was performed at a pressure of 60~Torr at temperatures of either 550~$ ^{\circ} $C or 600~$ ^{\circ} $C and V/III ratios between 70 and 150. Typical growth times ranged from 9~min to 11~min. The dynamics of InAs epitaxy in hybrid-TASE templates are described in the \href{}{Supporting Information}.\\
\textbf{Device contacting and gates patterning.}
After InAs growth, we spun a double layer of PMMA 669.04 (300~nm) and AR-P 672.03 (100~nm) resist and patterned device contacts with electron beam lithography. After resist development in methyl isobutyl ketone (MIBK) and isopropanol (IPA) with ratio 1:2, we etched the $ \sio $ template with BHF in exposed regions and passivated the InAs surface by immersion in 2~\% ammonium sulfide solution prior to evaporation of Ti (10~nm) and Au (150~nm). After lift-off in dimethyl sulfoxide (DMSO), we spun a single layer of AR-P 672.03 (100~nm) and patterned gate structures via electron beam lithography. The resist was developed in MIBK:IPA (1:2) and layers of Ti (5~nm) and Au (20~nm) were evaporated prior to lift-off in DMSO. After etching of native $ \sio $ in BHF, we metallized the Si handle wafer by evaporation of Ti (5~nm) and Pt (40~nm). During these steps, devices on the chip were protected by a 6.2~$ \mu $m thick layer of AZ~4562 optical resist.

\section{Acknowledgments}

This research was funded by DARPA (Grant No. 140D6318C0028). We thank W.~Riess and K.~Moselund for fruitful discussions and C.~Ciaccia, Z.~Lei, T.~Ihn and K.~Ensslin for help at early stages of the project. We thank D.~Caimi, S.~Paredes, M.~Jurich, A.~Bowers, the Cleanroom Operations Team of the Binnig and Rohrer Nanotechnology Center (BRNC) and especially A.~Olziersky, U.~Drechsler and S.~Reidt for their help and support. H.~Schmid and P.~Staudinger acknowledge funding from the European Union H2020 program SiLAS (Grant No. 735008). F.~Nichele acknowledges support from the European Research Commission (Grant No. 804273). The views, opinions and/or findings expressed are those of the author and should not be interpreted as representing the official views or policies of the Department of Defense or the U.S. Government.

\bibliography{libraryA.bib, combinedMARNotes.bib}



\clearpage
\newpage
\newcounter{myc} 
\renewcommand{\thefigure}{S.\arabic{myc}}

\section{Supporting Information}

\subsection{Supporting Information 1: Hybrid-TASE epitaxy along low symmetry directions}
\begin{figure}[H]
	\includegraphics[width=\columnwidth]{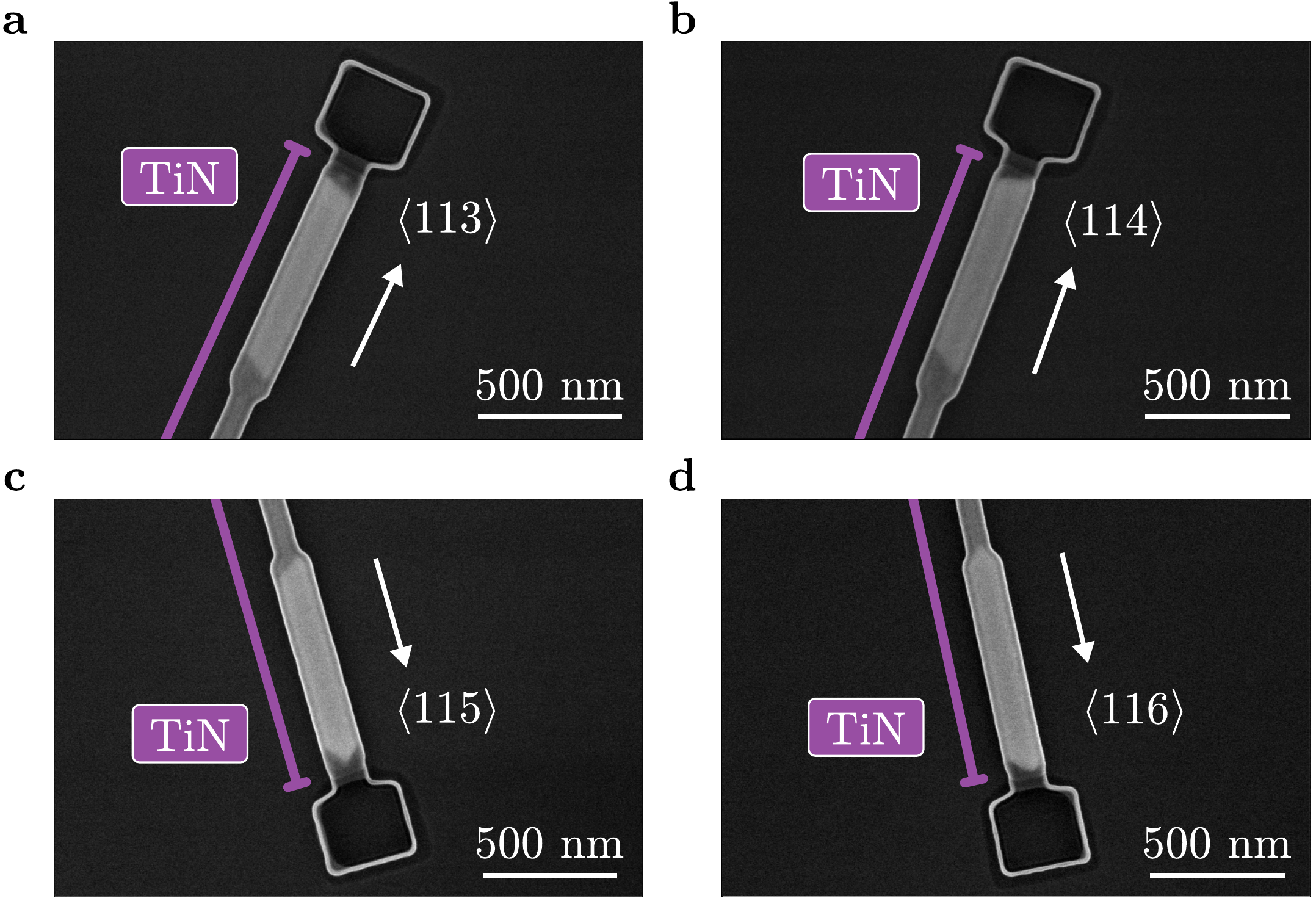}
	\caption{\textbf{Hybrid-TASE InAs epitaxy in lower symmetry directions.} Templates have similar aspect ratio and equal growth conditions as devices shown in Fig.~3 of the Main Text but different crystallographic alignment in \textbf{a} \dir{113}, \textbf{b} \dir{114}, \textbf{c} \dir{115} and \textbf{d} \dir{116} direction, respectively. Purple lines indicate the extent of TiN stripes inside hybrid templates.}
	\label{fig:S1}
\end{figure}

Uniform III-V semiconductor growth inside hybrid-TASE templates in multiple crystallographic directions can be of great importance for applications that require engineering of spin-orbit interaction \cite{Sasaki2014, Manchon2015, Carballido2021}. In the Main Text (Fig.~3e-h) we demonstrated that an isotropic growth rate in high-symmetry directions was accomplished via growth at As rich atmosphere with V/III=150.
In Fig.~\ref{fig:S1} we demonstrate growth in the lower symmetry \dir{113}, \dir{114}, \dir{115} and \dir{116} directions. The same results were obtained for templates oriented in \dir{115}, \dir{116}, \dir{221}, \dir{331} and \dir{441} direction (not shown). As discussed below, these results are not only a consequence of the global growth conditions in the reactor but also due to the presence of an exposed TiN surface inside the template cavities.

\setcounter{myc}{1}

\subsection{Supporting Information 2: InAs epitaxy dynamics in hybrid-TASE templates}
InAs epitaxy inside hybrid-TASE templates exhibited significantly altered growth dynamics compared to TASE epitaxy inside full $ \sio $ templates. Our observations include a change of facet morphology and increased axial growth rate. In the following, we investigate these effects at a range of epitaxy conditions and draw a direct comparison to TASE reference growth.\\

\begin{figure}
	\includegraphics[width=\columnwidth]{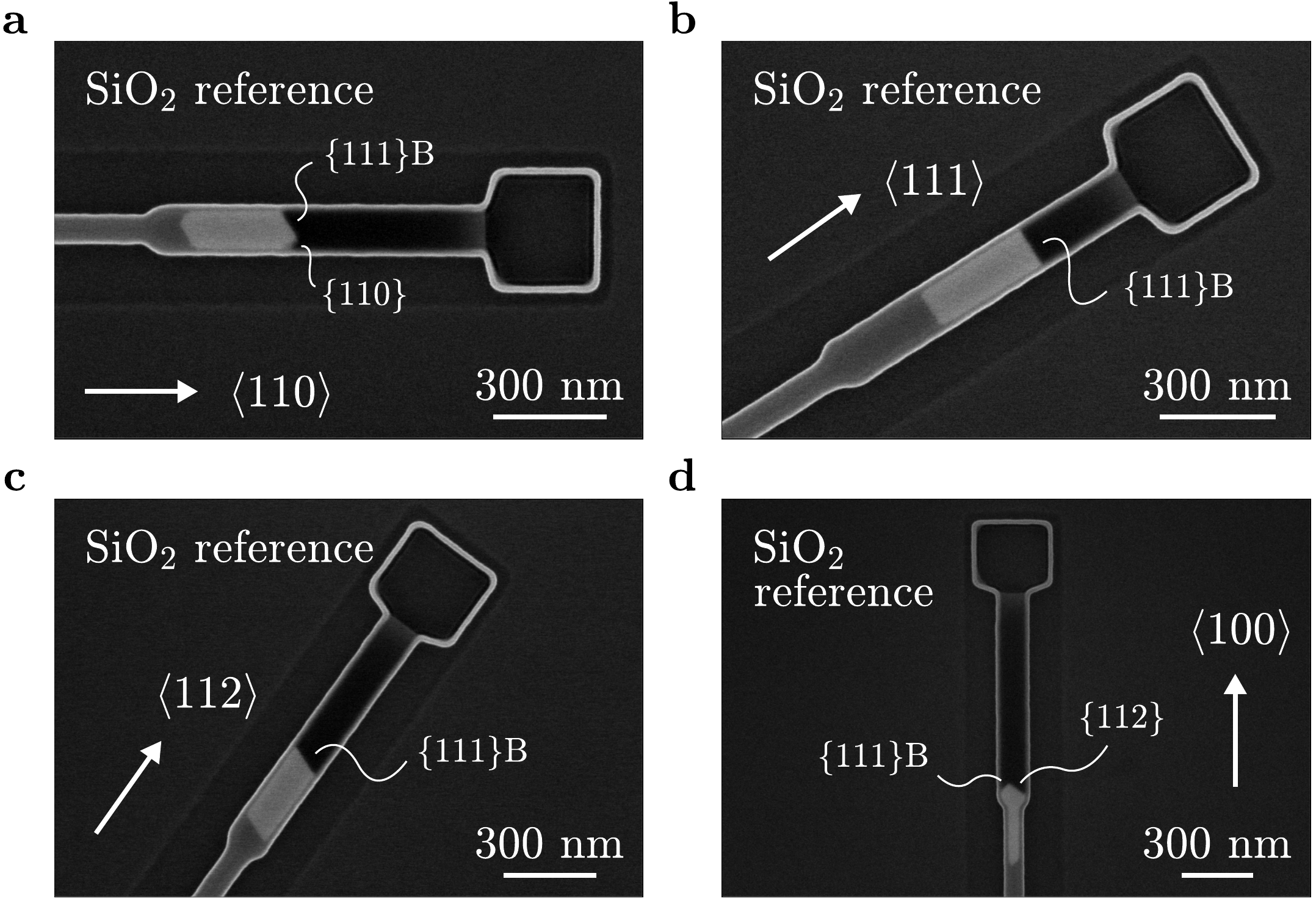}
	\caption{\textbf{TASE reference epitaxy.} InAs growth inside TASE templates without integrated TiN surface. Reference templates have a similar aspect ratio and were exposed to equal growth conditions as hybrid-TASE templates presented in Fig.~3 of the Main Text.  Crystallographic alignment of normal reference templates in \textbf{a} \dir{110}, \textbf{b} \dir{111}, \textbf{c} \dir{112} and \textbf{d} \dir{100} direction is equal to hybrid-TASE devices in the Main Text.}
	\label{fig:S2}
\end{figure}

To study the impact of exposed TiN inside the templates, we patterned $ \sio $ reference templates on the same chips as hybrid-TASE templates, thereby ruling out parameter variations between epitaxy runs as the origin of our observations. To pattern TASE cavities, we extended the area of TiN etching in Fig.~1c of the Main Text to remove TiN on the entire NW length.\\
Growth inside normal reference templates is presented in Fig.~\ref{fig:S2}. The reference templates have similar aspect ratio and equal crystal orientation as hybrid templates shown in Fig.~3 of the Main Text.  While the uniformity of growth rates along \dir{110}, \dir{111}, \dir{112} and \dir{100} template directions was reproduced in normal templates, the axial NW growth rate was reduced by more than a factor of two compared to hybrid epitaxy. \fac{110}, \fac{111}B and \fac{112} facets at the InAs growth front can be discerned in Fig.~\ref{fig:S2}, viewed through the thin $ \sio $ template layer. These results are consistent with faceting observed in previous studies on epitaxy in normal $ \sio $ templates \cite{Borg2014, Knoedler2017}. The high growth uniformity and radial filling of the templates as well as the formation of prominent \fac{111}B facets at As rich epitaxy has been attributed to enhanced growth of \fac{110} facets and suppressed growth in \dir{111} direction due to As trimer formation on the \fac{111}B facet \cite{Hayakawa1991, Moll1998, Tomioka2007}. We note that this was particularly apparent in normal templates along the \dir{111} and \dir{112} direction (Fig.~\ref{fig:S2}b and c), where we observe formation of single \fac{111}B facets, indicating that \fac{110} facets are grown out completely, despite the growth front is still deep inside the template.

\setcounter{myc}{2}

\begin{figure}
	\includegraphics[width=\columnwidth]{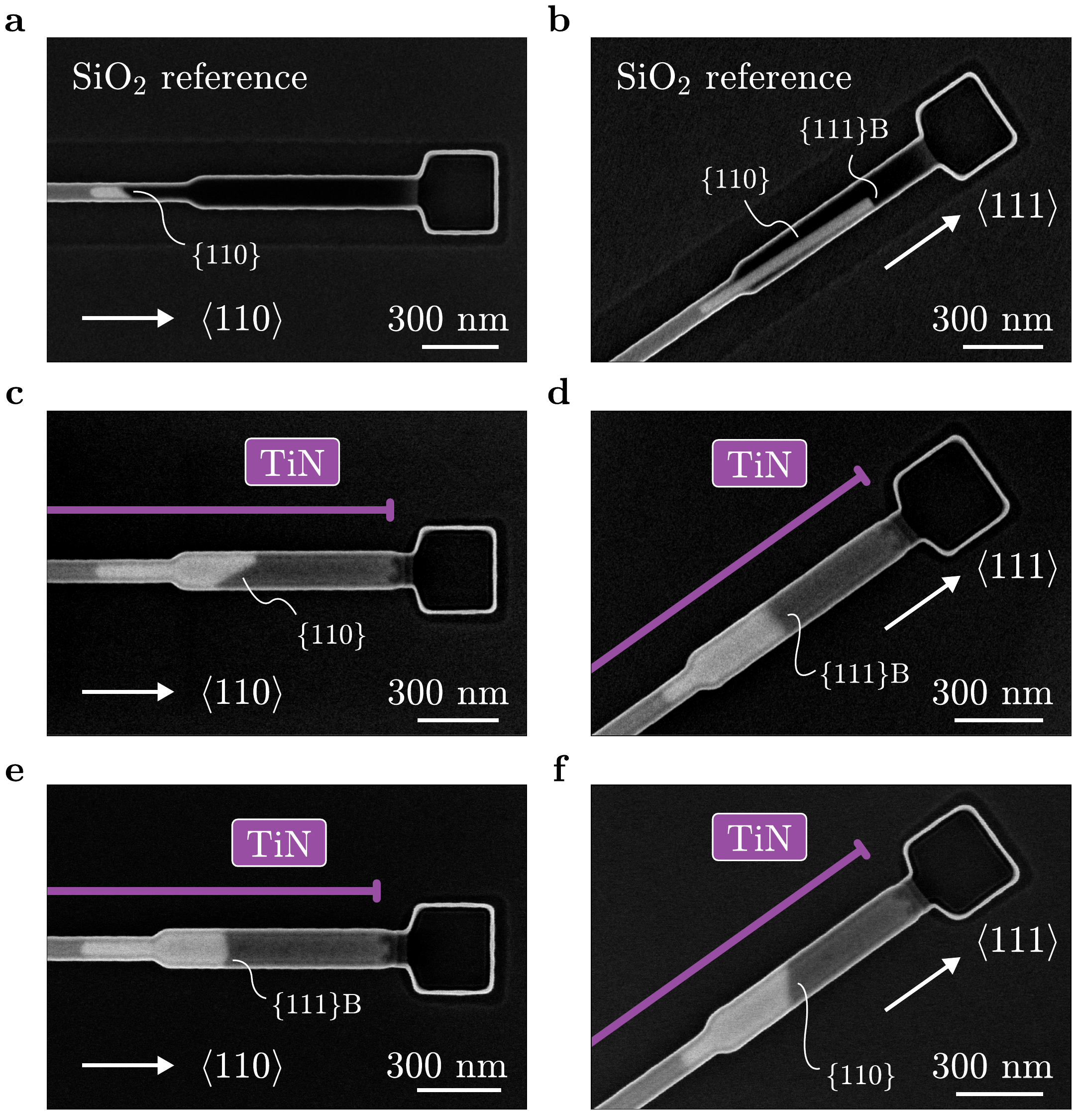}
	\caption{\textbf{Comparison of TASE and hybrid-TASE InAs epitaxy at low As flow.} InAs epitaxy with reduced precursor flow and V/III=70 inside TASE templates aligned along the \textbf{a} \dir{110} and \textbf{b} \dir{111} direction and inside hybrid-TASE templates along the \textbf{c, e} \dir{110} and \textbf{d, f} \dir{111} direction, respectively. Purple lines indicate the extent of TiN stripes integrated into hybrid templates.}
	\label{fig:S3}
\end{figure}

Results presented in Fig.~3 of the Main Text, and Figs.~\ref{fig:S1} and \ref{fig:S2} were obtained by growth in As rich atmosphere with V/III=150. Reducing the V/III precursor ratio to 70 and decreasing the TBAs and TMIn flows to $ 137~\mu\textrm{mol/min} $ and $ 1.9~\mu\textrm{mol/min} $, respectively, allowed us to study the dependence of the apparent growth rate enhancement inside hybrid-TASE templates as a function of crystal direction. Under these conditions, a growth time of 42~min was required to obtain devices presented in Fig.~\ref{fig:S3}. Importantly, the effective V/III ratio inside the templates was further decreased as a consequence of the high aspect-ratio of our template geometry \cite{Borg2015}. Hence, the new growth conditions resulted in an As sparse atmosphere. In Fig.~\ref{fig:S3} we compare epitaxy along the \dir{110} and \dir{111} direction inside normal and hybrid templates. When grown at low As flux, InAs does not cover exposed Si areas after nucleation \cite{Kanungo2013, Bjork2012}. To prevent the size of the Si seed to impact our comparison between TASE and hybrid-TASE growth in Fig.~\ref{fig:S3}, we etched cavities deep enough as to reach a constriction with lithographically defined cross section of 40~nm $ \times $ 70~nm.  The constriction smoothly expanded into a cavity with cross section 100~nm $ \times $ 70~nm, as the InAs growth front progressed. \\
In Figure~\ref{fig:S3}a we present InAs growth inside a normal template aligned along the \dir{110} direction. In this geometry we consistently obtained short wires that did not extend beyond the seed constriction but exhibited distinct \fac{110} facets. This behavior is consistent with growth at low V/III ratio, where growth of the \fac{111}B facet far exceeded that of the \fac{110} facet. For growth inside normal \dir{111} aligned templates (Fig.~\ref{fig:S3}b) this behavior was further emphasized as growth of the \fac{111}B facet was not limited by the template side walls. This resulted in a 770~nm long wire with a single \fac{111}B growth facet. Growth of the six \fac{110} NW side facets was suppressed completely and the NW diameter was effectively determined by the Si seed cross section and not by the template. The suppression of \fac{110} side facet growth corroborates the lack of group-V precursor molecules \cite{Jensen2004, Ikejiri2007}, since they access the template predominantly via Knudsen diffusion \cite{Borg2015, Borg2019} and are then incorporated at the \fac{111}B facet.

\setcounter{myc}{3}

The situation changed when a TiN surface was present in the cavity. In the \dir{110} aligned hybrid template (Fig.~\ref{fig:S3}c) the InAs crystal extended into the wide cavity region and radially expanded to the template walls. The increase in both radial and axial growth rate points towards a faster growth of \fac{110} facets compared to the reference growth in $ \sio $ templates. While the wire exhibited a \fac{110} facet as in the reference example, there was a second, most likely \fac{111}B facet, that is not grown out entirely, indicating a more balanced growth between \fac{110} and \fac{111}B facets. Again, our findings are confirmed by epitaxy in \dir{111} direction (Fig.~\ref{fig:S3}d) where a single \fac{111}B facet formed, similar to the reference (Fig.~\ref{fig:S3}b). However, the cavity was radially filled similar to growth in \dir{110} direction (Fig.~\ref{fig:S3}c), confirming faster growth of \fac{110} facets when a TiN surface was present. The NW in Fig.~\ref{fig:S3}d was 260~nm shorter than the reference, since the higher growth rate of \fac{110} facets caused a competition between \fac{110} and \fac{111}B facets for precursor material. We emphasize that such a decrease in axial growth rate was observed only in this specific situation, that is at reduced precursor flow in \dir{111} oriented templates. In all other growth regimes investigated in this work, accelerated growth in \dir{110} direction resulted in longer wires.\\
While growth in normal templates with the conditions of Fig.~\ref{fig:S3} always resulted in the growth morphology presented in Fig.~\ref{fig:S3}a and b, we observed a higher variability in faceting for the hybrid case. Two further examples are presented in Fig.~\ref{fig:S3}e and f where we observe at least one \fac{111}B facet in a \dir{110} hybrid template and a single \fac{110} facet in a \dir{111} hybrid template. A wide range of possible facet morphologies has previously been observed for GaAs and InAs epitaxy in TASE templates \cite{Borg2014, Knoedler2017}. The variability in morphology between individual wires has been ascribed to microscopic details of the Si seed morphology and nucleation on it. Consequently, the multitude of facet configurations in hybrid-TASE growth of Fig.~\ref{fig:S3}c-f points towards a more balanced growth equilibrium compared to epitaxy in reference templates of Fig.~\ref{fig:S3}a and b. In the latter case, the InAs facet morphology was entirely dominated by the As supply limited growth conditions.\\\\
Increased semiconductor growth rates have previously been reported in literature in the context of vapor-liquid-solid (VLS) NW growth and selective area growth (SAG) using a metal mask. Catalytic decomposition of group-III precursors molecules on the Au droplet surface caused a faster growth of GaP, GaAs and InAs VLS NWs \cite{Borgstrom2007} while fast lateral growth in SAG was attributed to enhanced surface diffusion of group-III precursors on a W mask surface \cite{Asai1985, Wernersson1995}. Growth selectivity in the latter process was found to significantly decrease for low growth temperature and rough metal surfaces \cite{Kobayashi2013, Soo2016}. We stress that in our work the smooth back face of an integrated TiN layer was exposed, the roughness of which was set by the native $ \sio $ layer below (see Fig.~3 of the Main Text). We speculate that the smooth TiN surface morphology lead to a more homogeneous surface energy that was less dominated by anisotropic surface energies of grain facets \cite{Patsalas2018}, enabling selectivity during hybrid-TASE epitaxy.\\\\
To summarize, fast epitaxy in hybrid-TASE templates could be the result of enhanced precursor surface diffusion on the integrated TiN segment. The change in facet morphology from TASE reference epitaxy to hybrid-TASE epitaxy (Fig.~\ref{fig:S3}) indicated an increased effective V/III material ratio at the growth front. The increase in \fac{110} facet growth rate leads us to the conclusion that the origin of the V/III ratio increase is a locally enhanced As vapor pressure rather than a reduced In pressure inside template cavities. We are not aware of previous work where an increased efficiency of group-V precursor incorporation by means of a metal surface was reported. Future work will include a detailed study of growth dynamics in the presence of metal surfaces.

\subsection{Supporting Information 3: Hybrid-TASE interfaces}
\begin{figure}
	\includegraphics[width=\columnwidth]{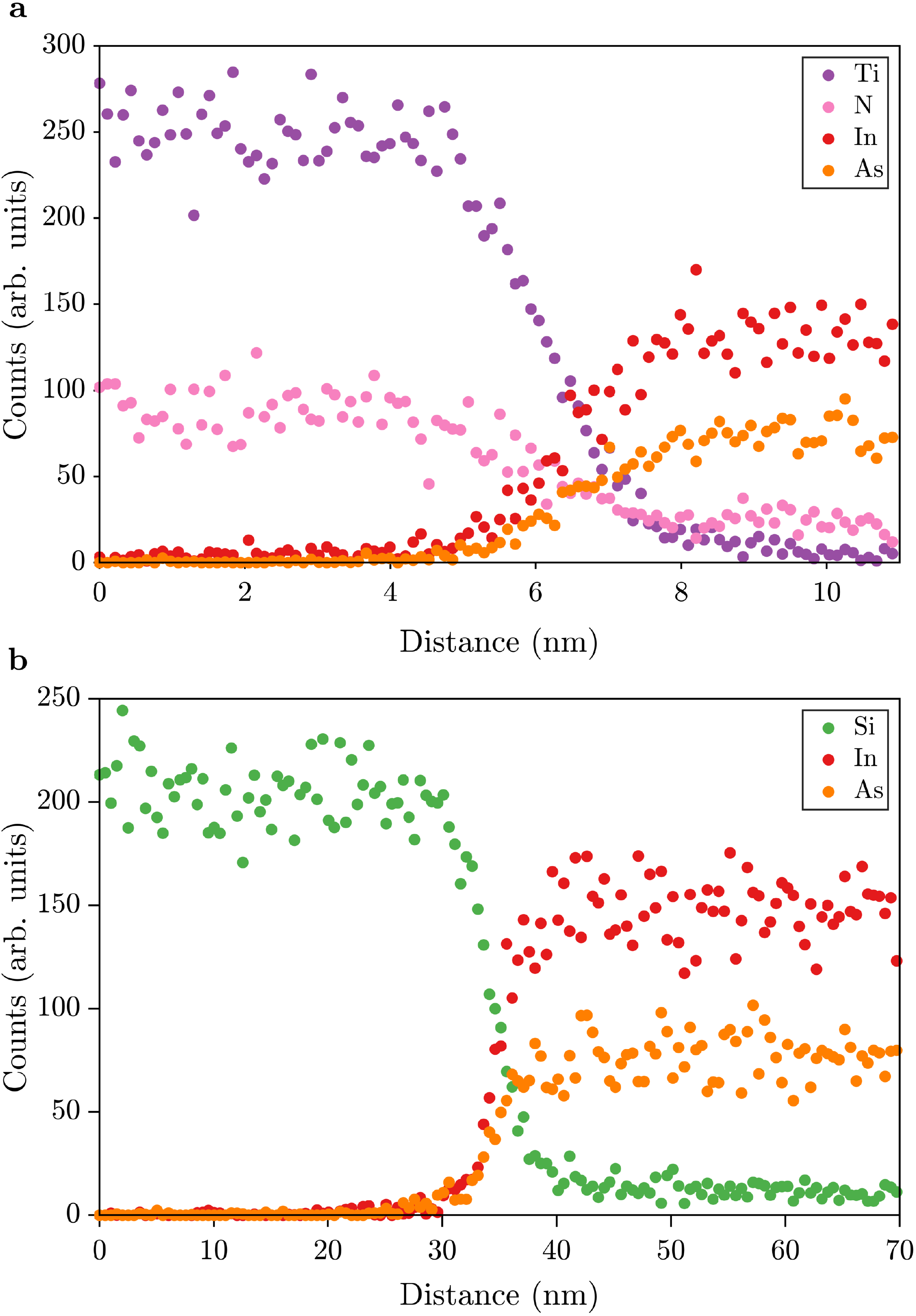}
	\caption{\textbf{EDX elemental line profiles.} \textbf{a} EDX line scan of the hybrid TiN/InAs interface of Fig.~2e of the Main Text. \textbf{b} EDX line scan of the Si/InAs seed interface of Fig.~3a of the Main Text.}
	\label{fig:S4}
\end{figure}

The characteristics of the hybrid TiN/InAs interface and the Si/InAs seed interface were discussed in the Main Text. Figures~\ref{fig:S4}a and b show EDX lines profiles of the interfaces presented in Figs.~2e and 3a of the Main Text, respectively.

\setcounter{myc}{4}

\subsection{Supporting Information 4: Electrical tuning of the tunnel junction}

\begin{figure}
	\includegraphics[width=\columnwidth]{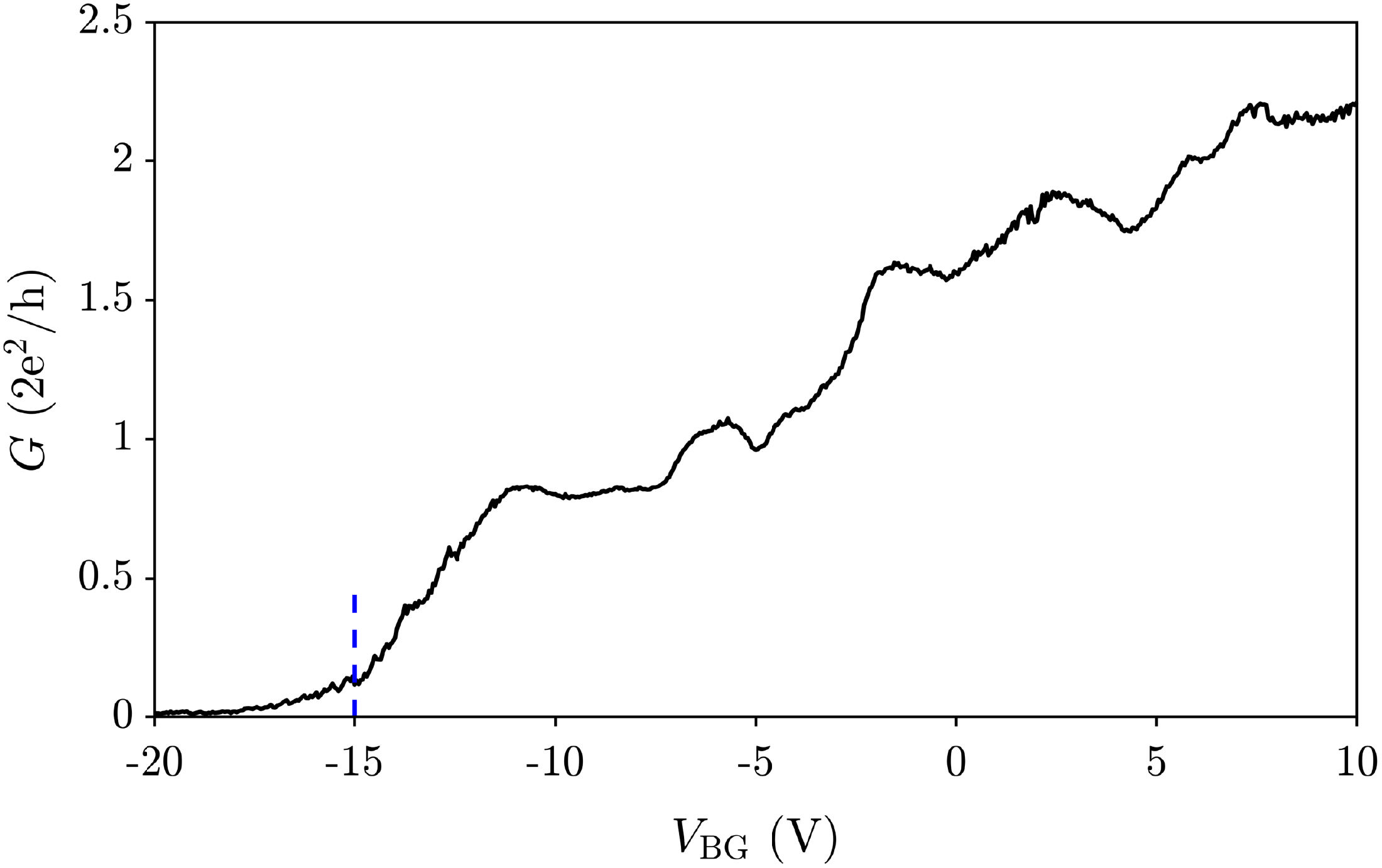}
	\caption{\textbf{Back gate tuning of a hybrid-TASE tunnel junction.} Tuning of a hybrid-TASE nanowire using the Si back gate at $ T=14~\textrm{K} $. Blue dashed line $ (V_{\textrm{BG}}  = -15~\textrm{V})$ indicates the operating point of measurements in Fig.~4 of the Main Text.}
	\label{fig:S5}
\end{figure}

At a temperature $ T=14~\mathrm{K} $, TiN in the device of Fig.~4a of the Main Text was resistive and we operated the metallized silicon handle wafer as a back gate. A representative pinch-off curve is presented in Fig.~\ref{fig:S5}. Oscillations in the conductance trace were reproduced throughout several sweeps of the back gate voltage $ V_\textsc{BG} $, consistent with conductance fluctuations in the wire. Measurements in Fig.~3 of the Main Text were performed at $ V_\textsc{BG} =-15~\textrm{V}$, indicated by the dashed line in Fig.~\ref{fig:S5}.

\setcounter{myc}{5}

\end{document}